# Keyword Optimization in Sponsored Search Advertising: A Multi-Level Computational Framework

Yanwu Yang[1], Bernard J. Jansen[2], Yinghui Yang[3], Xunhua Guo[4], Daniel Zeng[5]

[1]School of Management, Huazhong University of Science and Technology, Wuhan, China

[2]Qatar Computing Research Institute, HBKU, Doha, Qatar

[3]Graduate School of Management, University of California, Davis, CA, USA

[4]School of Economics and Management, Tsinghua University, Beijing, China

[5]Institute of Automation, Chinese Academy of Sciences, Beijing, China

**Abstract:** In sponsored search advertising, keywords serve as an essential bridge linking advertisers, search users and search engines. Advertisers have to deal with a series of keyword decisions throughout the entire lifecycle of search advertising campaigns. This paper proposes a multi-level and closed-form computational framework for keyword optimization (MKOF) to support various keyword decisions. Based on this framework, we develop corresponding optimization strategies for keyword targeting, keyword assignment and keyword grouping at different levels (e.g., market, campaign and adgroup). With two real-world datasets obtained from past search advertising campaigns, we conduct computational experiments to evaluate our keyword optimization framework and instantiated strategies. Experimental results show that our method can approach the optimal solution in a steady way, and it outperforms two baseline keyword strategies commonly used in practice. The proposed MKOF framework also provides a valid experimental environment to implement and assess various keyword strategies in sponsored search advertising.

# 1. INTRODUCTION

We have witnessed the deepening integration of online marketing and search technologies (Yang et al., 2017). This development has led to a primary format of online advertising, i.e., sponsored search advertising. Millions of advertisers choose sponsored search advertising to promote their products and services, taking advantages of precise targeting, low advertising costs and high return on investment. Internet advertising revenues hit a record high of $88.0 billion in 2017, where search advertising accounts for 46% of that pie (IAB, 2017).

In sponsored search advertising, advertisers choose a set of keywords of interest, and organize these keywords according to advertising structures (e.g., campaign and adgroup) defined by search engines. Once a user submits a query associated with one or more of these keywords to a search engine, it triggers a search auction process. That is, the search engine retrieves its advertisement databases and obtains a set of matching advertisements, which is then often ranked via the product of advertisers' bids prices and quality score, and priced through an auction mechanism (e.g., Generalized Second Price) (Che et al., 2017). This process generates a set of top-n advertisements that are displayed on search engine result pages together with organic search results. Throughout this entire procedure, keywords play a vital role in connecting advertisers, search users and search engines. Therefore, how to select profitable keywords and effectively manipulate keyword portfolios is a critical issue for search advertisers.

In sponsored search advertising, current research efforts along the line of keyword strategies mainly focus on extracting keywords from web pages for advertisement targeting (Yih et al., 2006; Ravi et al., 2010), generating related but low volume and inexpensive keywords with a small set of seed keywords (Joshi and Motwani, 2006; Fuxman et al., 2008), and identifying the most profitable set of keywords (Rusmevichientong and Williamson, 2006; Kiritchenko and Jiline, 2008). Existing efforts deal with a single keyword decision separately, while ignoring advertising structures in sponsored search. None has systematically studied keyword decisions at different levels throughout the entire advertising lifecycle, which is the focus of this research.

This is a challenging problem to study. First, the total number of distinct search terms is enormous (Bartz, 2006); however, advertisers, especially those from small and medium enterprises, have limited marketing resources for search advertising campaigns (Shin, 2015). Second, keywords have a rich set of performance indexes (Hu et al., 2010), containing various types of signals such as brand (Rosso and Jansen, 2010), geo-space (Yang, 2012) and demographics (Jansen et al., 2013). The returns for keyword investment vary significantly (Li et al., 2016). This leads to overlaying effects among keywords and advertising campaigns (Rutz et al., 2013). Third, the search advertising environment is essentially dynamic (Yao and Mela, 2011; Yang et al., 2015). Fourth, throughout the entire lifecycle of search advertising campaigns, advertisers have to deal with various keyword decisions at different levels. Moreover, keyword management costs inhibit advertisers' participation and reduce search engines' profits (Amaldoss et al., 2015). While these challenges mount, it's crucial to develop an integrated framework for keyword optimization in sponsored search advertising, in order to maximize the payoff from advertising dollars.

In this research, we formulate various keyword decisions in sponsored search advertising as a chain of related optimization problems in a hierarchical programming framework. This work develops an integrated multi-level computational framework for keyword optimization (MKOF) supporting various keyword decisions throughout the lifecycle of sponsored search advertising campaigns. First, we analyze keyword decision scenarios existing in sponsored search advertising. Second, we build a multi-level keyword optimization framework (MKOF) with consideration of various keywords decisions during the entire lifecyle of search advertising campaigns. Third, we establish a simple but illustrative instantiation of the MKOF framework, taking into account relationships between keywords, which will be explored in detail in Section 3.2. In the rest of this paper, we use "keyword set" and "keyword graph" interchangeably. We also conduct computational experiments to validate our framework and identified strategies, with two real-world datasets obtained from field reports and logs of search advertising campaigns. Experimental results show that our method can approach the optimal solution in a steady way, and it outperforms two baseline strategies in terms of total payoff.

The remanding of this paper is organized as follows. Section 2 discusses keyword related decision scenarios and presents a multi-level keyword optimization framework (MKOF). Section 3 provides an instantiation of the MKOF framework. Section 4 reports experimental results to validate the MKOF framework and instantiated strategies. Section 5 discusses possible contributions and managerial implications from this research. Section 6 concludes this work.

## 2. Keyword Optimization for Search Advertising

### 2.1. Keyword Decision Scenarios

Advertisers face various keyword decisions throughout the entire lifecycle of advertising campaigns in sponsored search. During this process, many factors could affect these keyword decisions. In general, there are four different levels of keyword decision scenarios (Figure 1):

1) Domain-specific keyword pool generation: Within a domain or industry (e.g. apparel), an advertiser needs to build and maintain a pool of relevant keywords from which a set of keywords for her search advertising campaigns to promote certain products/services can be determined. The output for this step is a set of keywords (we call it the domain-specific keyword pool).

2) Keyword Targeting (the Market-level Keyword Optimization): When an advertiser wants to promote a product/service (or several products/services together) (e.g., boots) on a given search engine, she needs to select a more accurate set of keywords from the domain-specific keyword pool in order to better fit the certain products/services and a particular search engine. This process takes into account how the search engine works, product features, as well as the target consumer population. We call this process Keyword Targeting or Market-level Keyword Optimization. The output is a set of keywords called the Target Keyword Set.

3) Keyword Assignment and Grouping (the Campaign and Adgroup level Keyword Optimization): To promote certain products/services, an advertiser often needs to design several advertising campaigns, and each campaign also includes several adgroups (we call this the basic search advertising structure). Given that a set of keywords generated from Keyword Targeting (i.e., the Target Keyword Set) is determined, the advertiser have to assign a subset of keywords for each campaign, and each campaign-specific keyword set also needs to be grouped into several groups, one for each adgroup. The Campaign level Keyword Optimization is called Keyword Assignment, and the Adgroup level Keyword Optimization is called Keyword Grouping. The output is called the Keyword Structure.

4) Keyword Adjustment: In ongoing search advertising campaigns, an advertiser has to dynamically adjust the Keyword Structure according to the real-time advertising performance (e.g. Click-Through-Rate, Cost-Per-Click, Conversion-Rate, etc.).

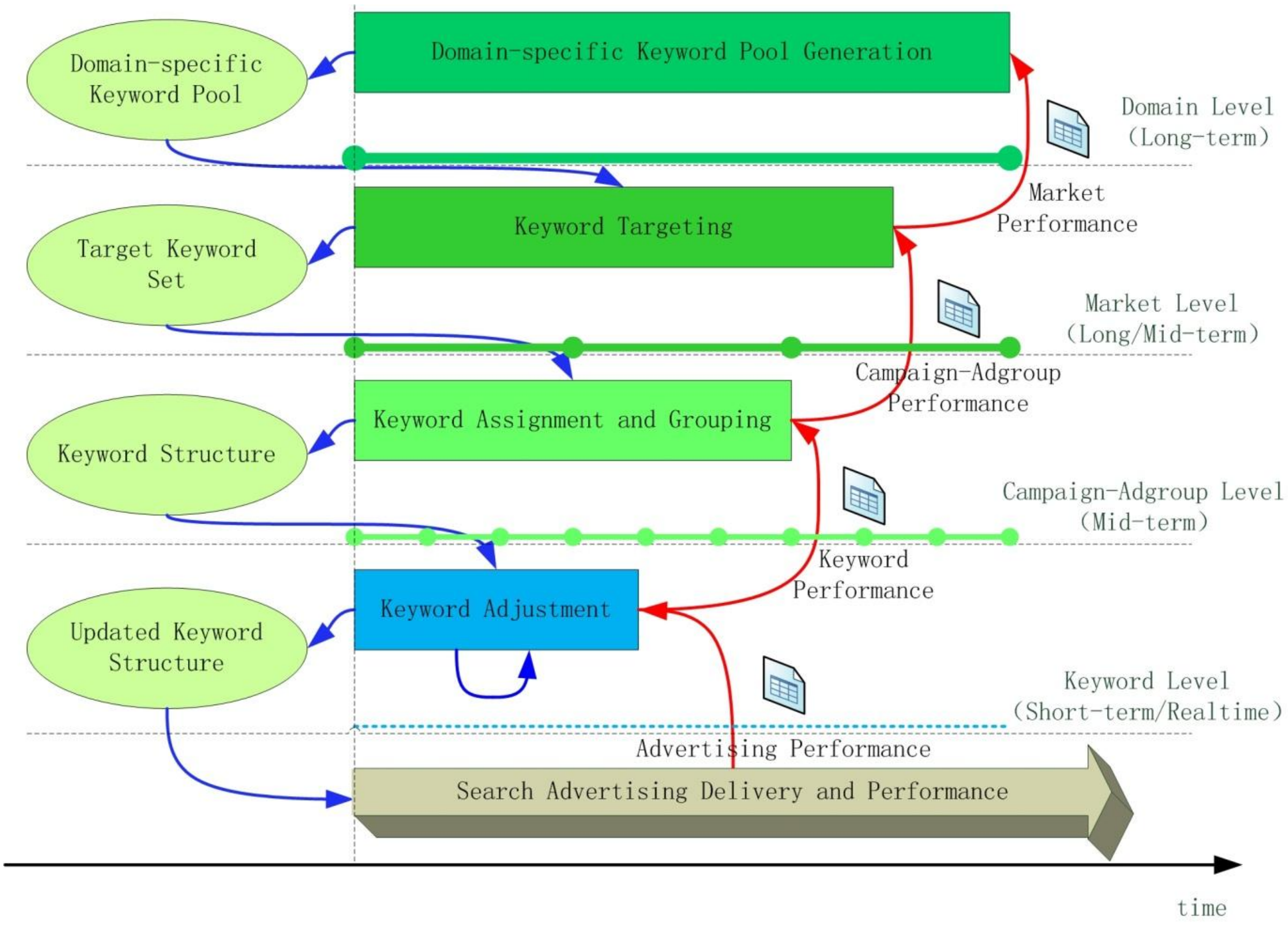

Figure 1. Keyword decisions in sponsored search advertising

In sponsored search advertising, keyword decisions at different levels form a closed-loop decision cycle. That is, results from higher-level decisions become constraints/inputs for lower-level decisions, and operational results at lower levels create feedbacks for decisions at higher levels. Thus, it is necessary to develop an integrated computational framework for advertisers to optimize various keyword decisions in sponsored search advertising.

### 2.2. A Keyword Optimization Framework

This research focuses on understanding keyword decision scenarios throughout the entire lifecycle of search advertising campaigns, and building a computational framework and optimization strategies at three levels: the market level (i.e., II), the campaign level and the adgroup level (i.e., III). In this research, we take the domain-specific keyword pool as given (the output of level I). Even though we do not perform detailed keyword adjustment (level IV) in this paper, we do calculate the advertising performance and keyword related performance from real-world advertising campaigns to provide feedbacks for decisions at higher levels.

This paper presents a Multi-level Keyword Optimization Framework (MKOF) with a hierarchical programming structure (Figure 2), with consideration of keyword decisions in the entire lifecycle of search advertising campaigns. In Figure 2, $U^*$ indicates the optimal solution of the keyword-level decision, which we do not study in this research. $V$ indicates the keyword-level feedback such as the keyword related performance, which we calculate from the real-world data. In the following we provide more explanations about other symbols used in this figure. In the MKOF framework, keyword decision at each level (e.g., market, campaign and adgroup levels) can be formulated as a mathematical programming model. The result of the higher-level model determines the input for the lower-level model, and the latter provides feedbacks to the former in the form of payoffs. Thus, the MKOF is a closed-loop, multi-level computational framework.

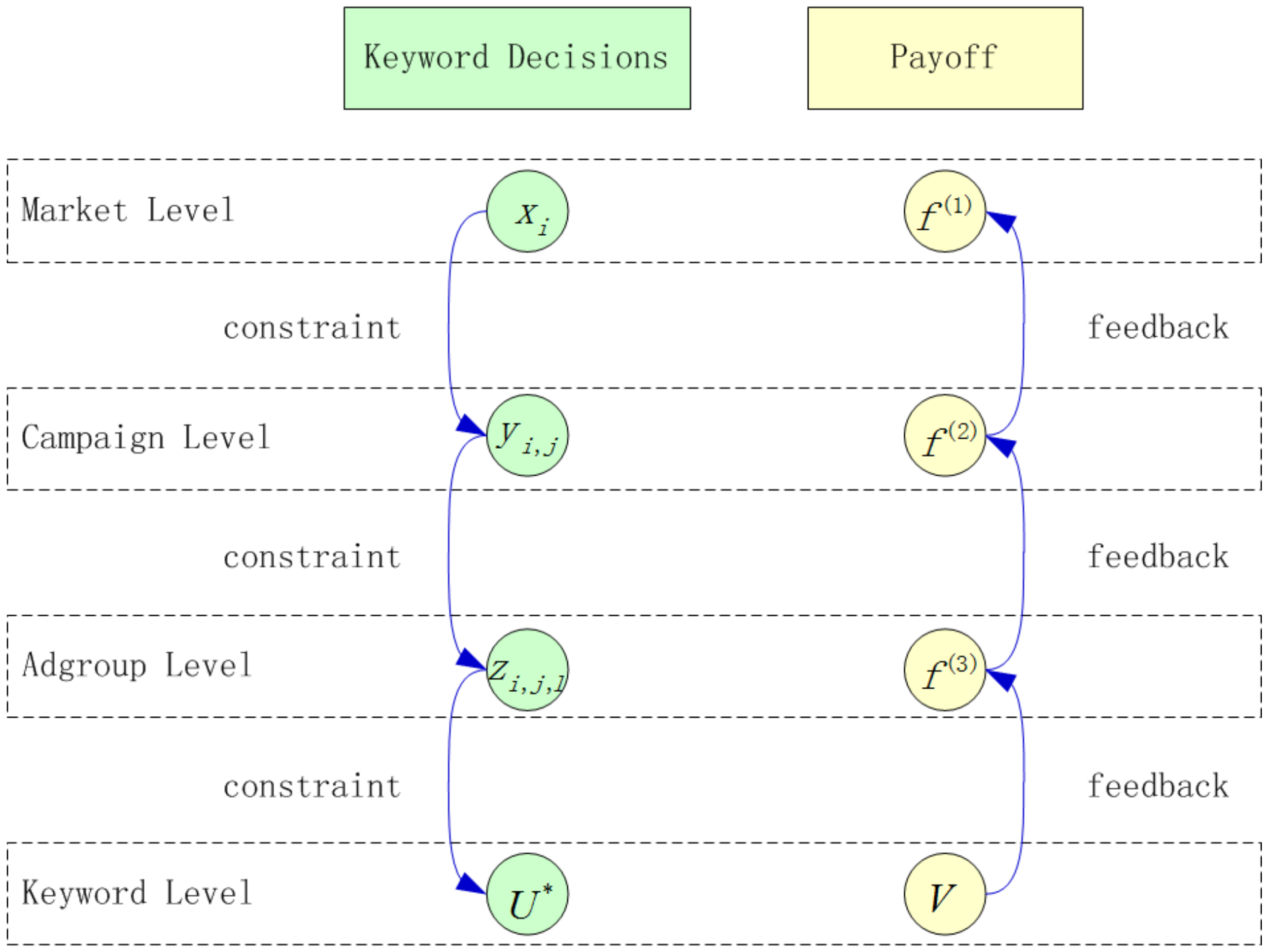

Figure 2. A multi-level computational framework for keyword optimization (MKOF) in sponsored search advertising

1) The Market Level: Let $\mathrm{K}^a$ denotes the domain-specific keyword pool for an advertiser $\mathrm{a}, \mathrm{a} = 1,2, \dots, \mathrm{r}$, who runs advertising campaigns across $n_1$ search markets, then the keyword selection at the market level (i.e., keyword targeting) can be given as,

$$\varphi: \mathrm{K}^a \to x_1^a, \dots, x_i^a, \dots, x_{n_1}^a, i \in \{1,2, \dots, n_1\},$$

where $x_i^a$ denotes the keyword set selected for a given search market $i$.

2) The Campaign Level: Keyword decision at the campaign level aims to select keywords for each campaign, for an advertiser, from the keyword set selected for a given search market, which can be given as,

$$\phi: x_i^a \to y_{i,1}^a, \dots, y_{i,j}^a, \dots, y_{i,n_2}^a, \mathrm{j} \in \{1, 2, \dots, n_2\},$$

where $y_{i,j}^a$ denotes the keyword set assigned to the j-th campaign in search market $i$. In the decision of keyword assignment, a keyword can be assigned to one or more campaigns, which is contingent on the relevance between keywords and characteristics of advertising campaigns.

3) The Adgroup Level: Keyword decision at the adgroup level aims to segment the keyword set of a given campaign into several subsets forming adgroups, which can be given as,

$$\psi:\ y^a_{i,j} \rightarrow z^a_{i,j,1}, \dots, z^a_{i,j,l}, \dots, z^a_{i,j,n_3}, l \in \{1, 2, \dots, n_3\},$$

where $z^a_{i,j,l}$ denotes the keyword set assigned to the $l$-th adgroup of the j-th campaign in search market $i$. In the search advertising structure, adgroup is the basic unit to encapsulate keywords within a certain campaign. Each ad group contains one or more ads and a shared set of keywords.

# 3. Mathematics of the Keyword Optimization Framework

## 3.1. Formulation

In this section, we present mathematics of the MKOF framework. In particular, following Yang et al. (2012), we formulate the MKOF framework as a three-level programming model with consideration of keyword decisions during the entire lifecycle of advertising campaigns in sponsored search.

**Model 1 (Market Level Model).** *Let $h^{(1)}$: $\mathbb{R}^{n_1+p} \rightarrow \mathbb{R}$ be the relevance function at the market level, $g^{(1)}$: $\mathbb{R}^{n_1} \rightarrow \mathbb{R}^{m_1}$ be constraints at the market level, $f^{(2)}$: $S_1 \rightarrow \mathbb{R}^p$ be the optimal relevance function at the campaign level, $S_1 \subset \mathbb{R}^{n_1}$, then the keyword optimization at the system level can be modeled as,*

$$f^{(1)} \coloneqq max_x\ h^{(1)}\left(x, f^{(2)}(x)\right)$$

$$s.t.\, g^{(1)}(x) \leq 0, \tag{1}$$

$$x \in X \subset \mathbb{R}^{n_1}.$$

**Model 2 (Campaign Level Model).** *Let $h_i^{(2)}$: $\mathbb{R}^{n_2+q} \rightarrow \mathbb{R}$ be the relevance function at the campaign level, $g_i^{(2)}$: $\mathbb{R}^{n_1+n_2} \rightarrow \mathbb{R}^{m_2}$ be constraints at the campaign level, $f_i^{(3)}$: $S_2 \rightarrow \mathbb{R}^q$ be the optimal relevance function at the adgroup level, where $i \in \{1, \dots, p\}$ and $x \in$* $S_1 = \{x \in X, g^{(1)}(x) \leq 0\}$, $S_2 \subset \mathbb{R}^{n_2}$, *then the keyword optimization at the campaign level can be modeled as,*

$$f_i^{(2)}(x) \coloneqq max_{y_i}\ h_i^{(2)}\left(y_i, f_i^{(3)}(y_i)\right)$$

$$s.t.\, g_i^{(2)}(x, y_i) \leq 0, \qquad (2)$$

$$y_i \in Y \subset \mathbb{R}^{n_2}.$$

**Model 3 (Adgroup Level Model).** *Let* $h_{i,j}^{(3)}\colon\ \mathbb{R}^{n_3+r} \rightarrow \mathbb{R}$ *be the relevance function at the adgroup level,* $g_{i,j}^{(3)}\colon\ \mathbb{R}^{n_2+n_3} \rightarrow \mathbb{R}^{m_3}$ *be constraints at the adgroup level,* $\hat{f}\colon\ S_3 \rightarrow \mathbb{R}^r$ *be the relevance function between keywords, adgroups and campaigns. where* $i \in \{1, \dots, p\}$, $j \in \{1, \dots, q\}$, $y \in S_2 = \{y \in Y, g^{(2)}(x, y) \leq 0\ \}$, *and* $z \in S_3 = \{z \in Z, g(x, y, z) \leq 0\ \}$ *where* $g(\cdot)$ *denotes constraints related to keywords, then the keyword optimization at the adgroup level can be modeled as,*

$$f_{i,j}^{(3)}(y) \coloneqq max_{z_{i,j}}\ h_{i,j}^{(3)}(z_{i,j}, \hat{f}(z_{i,j}))$$

$$s.t.\, g_i^{(3)}(y, z_{i,j}) \leq 0, \qquad (3)$$

$$z_{i,j} \in Z \subset \mathbb{R}^{n_3}.$$

### 3.2. Framework Instantiation

In this section, we establish a simple but illustrative model of keyword optimization within the mathematic framework proposed in Section 3.1.

In sponsored search advertising, an advertiser aims to obtain a set of frequent and relevant keywords for her products/services. Relevance is a kernel factor influencing advertising performance (Yang et al., 2018) and even the prospect of ecosystem of search advertising (Jansen, 2011). In particular, a higher relevance among keywords, advertisements and promoted products/services leads to a larger click-through rate and quality score[1] (Hillard et al., 2010). Moreover, wringing a high return from search advertising campaigns, as a matter of fact, is a matter of relevance[2]. Such that, relevant keywords should maximize the potential payoff the advertiser can obtain through the advertising campaigns. Therefore, we take the relevance as the criterion for various keyword decisions in sponsored search advertising. In keywords decisions, an advertiser aims to find a set of profitable keywords relevant to their products (or services). Thus, keyword decisions concern figuring out an optimal tradeoff between the relevance and the total profit.

[1] https://support.google.com/google-ads/answer/2454010?co=ADWORDS.IsAWNCustomer%3Dfalse&hl=en

Due to the incomplete information in the keyword targeting decision, we develop an iterative policy to approximate the optimum, i.e., a set of keywords maximizing both the relevance and profit across several search markets. This policy consists of two basic operations, namely an expansion and a trimming. Let $\pi_1$ denote the expansion operation that expands a graph based on keyword pool graph. For example, we first take the reference keyword graph, and expand it by finding the more related keywords on the keyword pool graph. Let $\pi_2$ be the trimming operation that eliminates unprofitable and less-profitable keywords from the expanded keyword graph. The expansion and trimming are done iteratively. After each round of iteration, we obtain a candidate Target Keyword Set, and then evaluate its quality (the weighted relevance and the payoff). Let $H_{\mathrm{k}-1}$ denote the reference keyword set ($H_0$ denotes seed keywords provided by the advertiser), $H_k$ be the Target Keyword Set, $H_e$ be a sub-graph of $H_{k-1}$ obtained using the trimming policy, and $\gamma$ be the revenue per unit budget, which can be obtained from the historical (or future) performance of ad-campaigns. Then $f^{(2)}(H_k)$ represents the weighted relevance of $H_k$ at the market level, and $x$ is the final version of $H_k$ at a given search market. The model of keyword optimization at the market level is given below.

$$
\begin{aligned}
f^{(1)} &:= \sum_{i=1}^{n_1} f_i^{(2)}(x) \\
s.t. \quad & \mathrm{H_e} = \pi_1(H_{k-1}) \\
& H_k = \pi_2(H_e, \gamma) \qquad\qquad (4) \\
& H_{k-1}, H_e, H_k \subset H \\
& \sum_{i=1}^{n_1} b_i - b \le 0
\end{aligned}
$$

where $\mathrm{f}_i^{(2)}$ is the maximum of the relevance score in the $i$th search market at the campaign

[2] https://www.wordstream.com/adwords-advertising

level, which is given as,

$$\mathrm{f}_i^{(2)}(x) \coloneqq \max \ \sum_{j=1}^{n_2} \mathrm{f}_{i,j}^{(3)}(y) \qquad (5)$$
$$\text{s.t.} \quad \sum_{j=1}^{n_2} b_{i,j} - b_i \leq 0$$

where $f_{i,j}^{(3)}$ is the maximum of the relevance score of the $j$th campaign in the ith search market at the adgroup level, which is given as,

$$f_{i,j}^{(3)}(y) := \max \hat{f}(z_{i,j}) + \hat{f}(z_{i,j,l})$$
$$\hat{f}(z_{i,j}) = \sum_{l}^{n_3} \hat{f}(z_{i,j}, z_{i,j,l}) \Big/ n_3 \qquad (6)$$
$$\hat{f}(z_{i,j,l}) = \sum_{k_a}^{z_{i,j,l}} \sum_{k_b}^{z_{i,j,l}} \hat{f}(k_a, k_b) \Big/ C_{|z_{i,j,l}|}^2$$

where $\hat{f}(z_{i,j})$ and $\hat{f}(z_{i,j,l})$ denote the relevance score between a given campaign and adgroups within it and that between keywords in a given adgroup, respectively. The former ensures that keywords assigned to a campaign fit its advertising theme, and the latter, postulates that keywords in a group has a good structure in terms of the relevance. For example, on one hand, a set of closed related keywords that are relevant to a campaign will get a higher value of $\hat{f}(z_{i,j})$, however, it may not be moderately divided into several groups; on the other hand, in the case with a set of keywords with manifest patterns of clusters, it thus results in a high value of $\hat{f}(z_{i,j,l})$, but not necessarily performs well if one or more clusters of keywords are not closely related to the campaign.

### 3.3. Solutions

In the following we provide solution algorithms for the MKOF framework and instantiated optimization models proposed in Section 3.2. Note that this research focuses on the potential benefits of keyword optimization framework with consideration of the basic advertising structure and a chain of various keyword related decisions during the entire lifecycle of advertising campaigns, rather than strategy optimization for an individual keyword decision.

Suppose that the Target Keyword Set is determined. For campaign $j$, a subset of

keywords is selected from the Target Keyword Set and forms a campaign-level keyword set $y_{i,j}$ (called Keyword Assignment). Generally, a keyword in the Target Keyword Set can be assigned to more than one campaign-level keyword set. This corresponds to a multiple choice knapsack problem (Babaioff et al., 2007), which aims to maximize the total relevance for each campaign, under a certain budget constraint (Yang et al., 2014). In particular, keywords are items to be selected, and each campaign can be considered as a knapsack. A campaign-level keyword set $y_{i,j}$ is then grouped into several keyword sets, each ($z_{i,j,l}$) corresponding to an adgroup in campaign $j$ (called Keyword Grouping). Generally, a keyword in $y_{i,j}$ can only belong to one adgroup $z_{i,j,l}$. Organizing closely related keywords with adgroups allows advertisers to reach the right consumers with her ad campaigns (Yang et al., 2018). For an advertiser, well-organized keyword groups improve ad group performance and her account's relevance, resulting in lower PPC charges (WordStream, 2018). The objective of Keyword Grouping is to maximize the relevance among keywords within an adgroup, thus can be formulated as a clustering problem. In keyword clustering, each keyword is represented as a vector of its neighbors. We implement a k-means clustering algorithm to segment keywords within a campaign. Algorithm 1 illustrates the procedure for solving the entire framework.

**Algorithm 1** (the solution for the entire MKOF framework)

**Input**: the domain-specific keyword pool graph, the seed keyword set, the number of campaigns, the budget constraint for each campaign, other necessary parameters derived from data (e.g. impressions, clicks, CPC, CTR, and revenue per click).

**Output**: the optimal keyword selection at different levels (i.e., market, campaign, and adgroup).

1) Expand and trim the reference keyword set $H_{k-1}$ (its initial set $H_0$ denotes seed keywords), and obtain the target keyword set $H_k$.

2) Solve the multiple-choice knapsack problem at the campaign level.

3) Solve the clustering problem at the adgroup level, and calculate the relevance score for each adgroup.

4) Calculate the relevance score at campaign level.

5) Calculate the relevance score at the market level.

6) Calculate the weighted relevance of $H_k$, and the incremental payoff. If $\delta > \varepsilon$ (i.e.., the incremental payoff is not small enough), let $H_{k-1}$= $H_k$, and go to step 1). Otherwise (if the terminating condition $\delta \leq \varepsilon$ is satisfied), output the final target keyword set at the market level (i.e., $x_i = H_k$), the assigned keyword set at the campaign level (i.e., $y_{i,j}$), and keyword grouping results (i.e., $z_{i,j,l}$).

Note that all calculations on keyword sets leverage relationships between keywords. In our model $H_{k-1}$ and $H_k$ are represented as keyword graphs, in which nodes represent individual keywords, and edges represent relationship between keywords. Relationships between keywords can be either semantic (e.g., is-a) or statistical (e.g., co-occurring). Relevance is calculated based on typical graph methods. In this research, we take the statistical co-occurring graph of keywords where an edge between two keywords represents whether they occur together in search users' queries (i.e., 1 if yes, 0 otherwise). The relevance between two keywords is computed as the inverse of the geodesic distance between them, i.e., $\hat{f}(k_a, k_b) = {}^{1}/_{d(k_a, k_b)}$.

In a nutshell, our model of keywords optimization operates in an iterative way, i.e., Keyword Targeting $(\pi_1 \rightarrow \pi_2)$ $\rightarrow$ Keyword Assignment $\rightarrow$ Keyword Grouping $\rightarrow$ Evaluation $(f_{i,j}^{(3)}(y) \rightarrow \mathrm{f}_i^{(2)}(x) \rightarrow f^{(1)})$ $\rightarrow \cdots$ When the incremental payoff is small enough, the entire procedure of keywords optimization terminates, and then we obtain the final Target Keyword Set and the Keyword Structure.

Although the solution for the entire MKOF framework is a simple greedy algorithm, it could steadily approach the optimum (or the semi-optimum). Since $H_e = \pi_1(H_{k-1})$, the number of keywords in $H_e$ is larger than that in $H_{k-1}$. Such that, $profit(H_e) \geq profit(H_{k-1})$. Moreover, in the operation of $H_k = \pi_2(H_e, \gamma)$, the keywords with zero or negative profits are removed, so $profit(H_k) \geq profit(H_e)$. After each iteration, we have $profit(H_k) \geq profit(H_{k-1})$, i.e., the profit is non-decremental. In other words, for any subset of $H_x$, we have $profit(H^*) \geq profit(H_x)$. Therefore, the iterative algorithm for the MKOF framework could generate a relative steady set of keywords ($H^*$) after a limited number of iterations. This property ensures that Algorithm 1 could converge to an optimal solution after a limited number of iterations.

# 4. Experimental Validation

### 4.1. Data Preparation and Experimental Setup

In this section, we conduct computational experiments to validate our framework and instantiated strategies for keyword decisions in sponsored search advertising, with two real-world datasets from field reports and logs capturing search advertising activities. The evaluation focuses on two-fold purposes. The first aims to evaluate the performance of our framework and instantiated keyword strategies, by comparing with two baseline strategies. Second, we report more details of these keyword decisions to validate some properties of our framework.

The two datasets includes 121 and 129 keywords, respectively. We take each of these two keyword sets as the domain-specific keyword pool in the following experiments. For each keyword, the dataset records the number of impressions, the number of clicks, cost-per-click, click-through rate. The revenue per click can be obtained from the statistics derived from past search advertising campaigns. The co-occurring statistics between keywords were obtained from Google Adwords's Keyword Planner or third-party tools (e.g., WordTracker). Then we can generate a binary undirected graph of keywords to represent a given domain-specific keyword pool.

The following experiments are set up as follows. Given that a domain-specific keyword pool, the reference keyword set provided by the advertiser, the number of campaigns, the budget constraint for each campaign, and the number of adgroups are determined, our keyword framework obtains the Target Keyword Set (at the market level) and the Keyword Structure (at the campaign and adgroup levels). In the experiment on the first dataset (Case-1), we take the setting where an advertising account on a search engine including three campaigns, and each campaign consists of two adgroups in order to make comparisons with results from the practical operations[3]. Likewise, in the experiment on the second dataset (Case-2), we consider another setting including two campaigns, and each campaign consists of two adgroups.

### 4.2. Comparisons

We compare our keyword optimization framework (MKOF) and instantiated strategies with two baseline strategies to validate the effectiveness of our method. For comparison purposes, we implemented two baseline strategies. The first baseline strategy is the one used by the firm in past advertising campaigns, called BASE1-Origin. It allocates the Target Keyword Set determined by our method into a given search advertising structure, in order to illustrate the performance of keyword assignment and grouping strategies. The second baseline strategy is a heuristic approach based on the ratio of CTR and CPC, called BASE2-Ratio. It selects top-n keywords according to the CTR-CPC ratio. It first selects top-n keywords from the domain-specific keyword pool to form the target keyword set, and then the target keyword set is assigned to campaigns by using our method (e.g., the multiple-choice knapsack and k-means clustering algorithm). BASE2-Ratio is based on a specific heuristic (i.e., the CTR-CPC ratio) for advertisers to maximize advertising efficiency, because CTR is one of major factors influencing the quality score in sponsored search advertising (Google Adwords, 2018). A higher quality score could results in more clicks with lower costs (Yang et al., 2015). On one hand, the comparison with BASE1-Origin is to evaluate the performance of our keyword assignment and grouping strategies, because both BASE1-Origin and our method use the same target keyword set. On the other hand, the comparison with BASE2-Ratio is to evaluate our keyword targeting strategy, because BASE2-Ratio and our method use different target keyword sets while keeping the rest procedure the same. The initial reference sets for Case-1 and Case-2 include three and two keywords covering companies' names and primary products, respectively. Table 1 presents the resulting performance of our MKOF and two baseline strategies.

| Strategy | Payoff | |
|---|---|---|
| | Case-1 | Case-2 |

[3] Note that determining the number of campaigns and the number of adgroups is beyond scope of this research, which might be covered in the future work.

| | | |
|---|---|---|
| MKOF | 11,938,036 | 61,585 |
| BASE2-Ratio | 10,387,413 | 57,992 |
| BASE1-Origin | 11,108,298 | 57,605 |

Table 1. Performance of the MKOF and two baseline strategies

From Table 1, we can see the following observations.

(1) In Case-1, the payoff of our MKOF method is 7.5% (14.9%) higher than that of the BASE1-Origin (the BASE2-Ratio). In Case-2, the payoff of our MKOF method is 6.9% (6.2%) higherer than that of the BASE1-Origin (the BASE2-Ratio). Thus, we can conclude that our MKOF method outperforms the two baselines strategies in terms of payoff.

(2) The superiority of our method over the BASE2-Ratio indicates that the keyword targeting strategy (at the market level) in our MKOF framework can effectively find a superior target keyword set, with a small reference keyword set as seeds.

(3) The keyword assignment (at the campaign level) and keyword grouping (at the adgroup level) process is capable of finding a better way (based on the comparison with BASE1-Origin) to assign and group the target keyword set into campaigns and adgroups.

### 4.3. Procedure and Properties

The second set of experiment provides more details on the procedure and properties of our MKOF method. (I) Keyword Targeting (at the market level): In Case-1, using Algorithm 1 to expand and trim the initial set of keywords, we obtain a target keyword set with 80 keywords after 11 iterations. In Case-2, we obtain a target set of 109 keywords after 6 iterations. (II) Keyword Assignment (at the campaign level): In Case-1, we allocate the target keyword set to three advertising campaigns: 25 for campaign-1, 36 for campaign-2, and 52 for campaign-2. In Case-2, we allocate the target keyword set to two advertising campaigns: 59 for campaign-1 and 71 for campaign-2. (III) Keyword Grouping (at the adgroup level): In Case-1, we obtain the optimal keyword grouping result: campaign-1 includes two keyword groups (i.e., 7 and 18 keywords, respectively), campaign-2 has two keyword groups with 27 and 9 keywords, respectively, and campaign-3 has two keyword groups with 41 and 11 keywords, respectively. In Case-2, we obtain the optimal keyword grouping result:

campaign-1 includes two keyword groups (i.e., 41 and 18 keywords, respectively), and campaign-2 has two keyword groups with 49 and 22 keywords, respectively.

Figure 3 illustrates the evolution of the total payoff during the iterative procedure of our MKOF framework in the two experiments, respectively. From Figure 3, we can observe the following phenomena.

(1) In both cases, the total payoff generated by our MKOF method increases steadily as the iteration progresses, and eventually approaches a semi-optimal level within a finite number of iterations. This phenomenon confirms the fact that our MKOF method could converge quickly, resulting in a set of keywords with an optimal tradeoff between the total relevance and the total profit on the semi-optimal level.

(2) During the iterative procedure, the incremental payoff shows a decreasing trend, in both cases. The possible explanation is that, as the target set expands over the keyword graph, an additional unit of advertising budget can result in a less return. This can be justified by the law of diminishing marginal utility in economics (Mankiw, 1998).

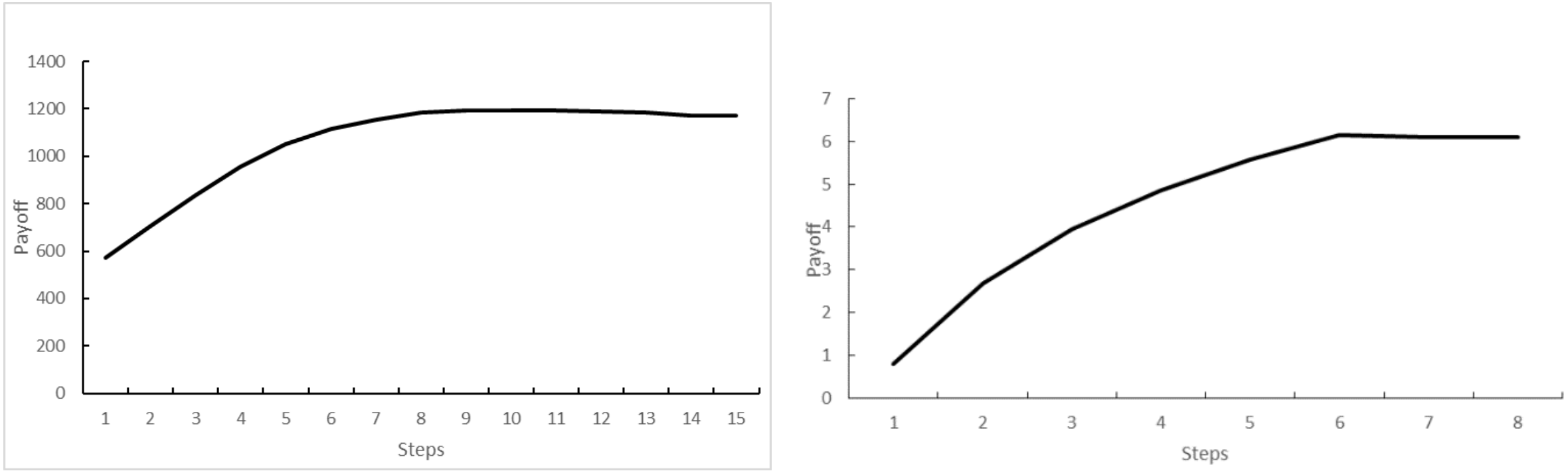

a) Case-1 b) Case-2

Figure 3. Evolution of the total payoff

## 5. Discussions

In summary, experimental results demonstrate that our proposed computational framework is a feasible solution to keyword decisions in sponsored search advertising. It outperforms two baseline strategies in terms of the total payoff. Moreover, our proposed computation framework provides a valid experimental environment and a closed-loop decision chain capable of implementing various keyword strategies during the entire lifecycle of advertising campaigns in sponsored search.

In addition to the methodology contribution, this research provides important managerial insights for advertisers in sponsored search advertising. First, in a novel advertising media such as sponsored search advertising, it's necessary for advertisers to find the right target keyword set covering the target market, rather than only focusing on each individual keyword's performance indexes. Second, search engines define various advertising structures and rules to facilitate online processes. The advertising performance might be severely weakened if advertisers ignore advertising structures while making keyword decisions. Third, it demands a systematic view to understand and make various keyword decisions throughout the entire lifecycle of advertising campaigns in sponsored search.

## 6. Conclusions and Future Directions

In this paper, we propose a novel hierarchical computational framework for keyword optimization in sponsored search advertising, with consideration of the entire lifecycle of advertising campaigns. We formulate the MKOF framework as a chain of optimization models, and provide a set of solution algorithms. We conduct computational experiments to validate its effectiveness with two real-world datasets of search advertising campaigns. Experimental results show that our method can approximate the optimal solution in a steady way, and it outperforms two baseline strategies in terms of the total payoff.

We are in the process of extending our framework and strategies in the following directions: (a) uncertainties and dynamics of keyword strategies in the MKOF framework; (b) effective algorithms specified for various keyword related decisions, including keyword targeting, keyword assignment, keyword grouping and realtime keyword adjustment; (c) advertising competition and game-theoretical approaches for keyword decisions (Yang et al., 2016); (d) co-optimization with other advertising strategies (e.g. bidding) in sponsored search.

Yanwu Yang is a full professor in the School of Management, Huazhong University of Science and Technology, and head of the ISEC research group (Internet Sciences and Economic Computing). His research interests include computational advertising, advertising decisions, web personalization, and user modeling. Yang has a PhD in computer science from the graduate school of école Nationale Supérieure d'Arts et Métiers. Contact him at yangyanwu.isec@gmail.com.

Bernard J. Jansen is a principal scientist in the social computing group at the Qatar Computing Research Institute, a professor in the College of Science and Engineering at Hamad bin Khalifa University, and a professor in the College of Information Sciences and Technology at Pennsylvania State University. Jansen has a PhD in computer science from Texas A&M University. He has served as a Senior Fellow at the Pew Research Center with the Pew Internet and American Life Project and a university expert with the National Ground Intelligence Center. Jansen is editor in chief of Information Processing & Management. Contact him at jjansen@ist.psu.edu.

Yinghui (Catherine) Yang is an associate professor of the Graduate School of Management at the University of California, Davis. Her research interests include business intelligence, big data and data mining, recommendation, online shopping, and service marketing. Yang has a PhD in operations and information management from the Wharton School at the University of Pennsylvania. Contact her at yiyang @ucdavis.edu.

Xunhua Guo is an associate professor of Information Systems at the School of Economics and Management, Tsinghua University. His research takes behavioral and design science approaches to topics on electronic commerce, social networks, and business intelligence. He has a PhD in information systems from Tsinghua University in 2005. Contact her at guoxh@sem.tsinghua.edu.cn.

Daniel Zeng is Gentile Family Professor in the Department of Management Information Systems at the University of Arizona, and a Visiting Research Fellow at the Chinese Academy of Sciences. His research interests include intelligence and security informatics, infectious disease informatics, social computing, recommender systems, software agents, spatial-temporal data analysis, and business analytics. He has a PhD in industrial administration from Carnegie Mellon University. He has published one monograph and more than 330 peer-reviewed articles. He served as the editor-in-chief of IEEE Intelligent Systems from 2013-2016 and currently serves as President of the IEEE Intelligent Transportation Systems Society. Contact her at zeng@email.arizona.edu.